\documentclass{aa}  

\usepackage{graphicx}
\usepackage{txfonts}
\usepackage{natbib}
\usepackage{makecell}
\usepackage{gensymb}
\usepackage{xcolor}
\usepackage{lipsum}
\usepackage{wallpaper}
\begin{document}

\title{Planet--moon ejections in close stellar encounters}


\author{Y. Badoux
      \inst{1,2}
      \and
      S. Portegies Zwart
      \inst{1}
      }

\institute{
$^1$ Leiden Observatory, University of Leiden, Einsteinweg 55, 2333 CC Leiden\\
$^2$ J{\"u}lich Supercomputing Center, Forschungszentrum J{\"u}lich, J{\"u}lich, German
         }

\date{Received 28 March 2026; accepted 10 August 2026}

\abstract
{Free-floating planet–moon pairs (FFPMs) may form when planetary systems experience close stellar encounters.}
{We aim to quantify moon-retention probabilities and ejection cross-sections for planets and planet–moon systems subject to close stellar encounters, and to identify the orbital conditions most favorable for forming FFPMs.}
{We conducted extensive numerical scattering experiments to compute ejection cross-sections on a grid in the orbital separation of the planet ($a_p$) and the moon ($a_m$), marginalizing over all other initial parameters. The simulations tracked both planetary and lunar fates across a wide range of encounter geometries.}
{FFPM cross-sections depend primarily on the moon's semi-major axis,
  $a_m$, decreasing gradually until a sharp drop near
  $a_m\sim0.45\,R_H$. Planet-only ejection cross-sections instead
  decline steadily with $a_p$. Moon survival remains near unity for
  Io-like orbits ($a_m = 0.008\,R_H$) and falls steeply beyond
  $0.4\,R_H$. Compared with planet–planet scattering, stellar
  encounters preserve moons more effectively across all
  separations. Surviving moons at $a_m\lesssim 0.4\,R_H$ retain
  near-circular, low-inclination orbits, while wide-orbit moons show
  stronger dynamical excitation. Applying our cross-section maps to
  the microlensing system MOA-2011-BLG-262Lb suggests possible
  progenitor semi-major axes between $a_p \sim 1.3$\,au and $\sim 5.9$
  au for a solar-mass host.  Even though uncertainties remain large
  and the probability distribution is rather flat, we prefer $a_p \sim
  5.2$\,au.}
{Stellar encounter ejections constitute a viable channel for producing FFPMs whose orbital and survival properties differ from those formed by planet–planet scattering. Moon retention and orbital excitation patterns provide promising diagnostics of ejection history. Current and upcoming microlensing and direct-imaging surveys may be capable of detecting Galilean-mass exomoons around rogue planets, offering new tests of dynamical formation pathways.}

\keywords{planets and satellites: dynamical evolution and stability; Astrophysics - Earth and Planetary Astrophysics}

\maketitle

\section{Introduction}

The discovery of free-floating planets (sometimes referred to as rogue
planets) has highlighted discussions concerning their formation.  Maybe
these planets formed in situ or are ejected when the host star
explodes in a supernova; but the common literature considers them to
be dynamically dissociated from their host star. There are two main
mechanisms that lead to the dynamical ejection of one or more planets:
(1) the internal instability of the planetary system in which one planet
kicks out another
\citep{1996Sci...274..954R,1996Natur.384..619W,1997ApJ...477..781L,2026ApJ...998..245H};
and (2) a passing star that leads to the ejection of one or more
planets in the encounter \citep{hills1984close}.  Both situations
naturally arise from the planet formation processes in a disk and
from the dynamical evolution of young stellar clusters.

We question whether we can make a distinction between these scenarios.  It seems difficult to do this based on the planet mass, size, or obliquity. However, since both ejection mechanisms have a dynamical
origin, maybe the presence of moons around rogue planets gives an
indication of its ejection mechanism.

Moons, so far, have not been observed orbiting rogue planets or around any extra-Solar-System planet. However, considering the copious presence of moons around giant planets in the Solar System, we consider them to be common, including around exoplanets.  We question under what
conditions a moon survives the ejection of its planet.  This has been
studied for planet--planet scattering, but not for planets ejected
through a close encounter with another star.

Several microlensing surveys have targeted free-floating planets: OGLE
\citep{OGLE}, MOA \citep{2008ExA....22...51S,2023AJ....166..108S}, and
KMTNet \citep{2016JKAS...49...37K}. Over the last two decades, these
surveys have detected the first free-floating planets. In 2014, OGLE
and MOA reported the first exomoon around a rogue planet candidate:
MOA-2011-BLG-262Lb \citep{MOA2011}. The light curve permits degenerate
interpretations: a sub-Earth-mass moon orbiting a Jupiter-mass planet
or a high-velocity planetary system.  The latter is supported by
\cite{2025AJ....169..131T}, who favor a $29 \pm 5$~M$_\oplus$
mass planet orbiting a $0.19 \pm 0.03$~M$_\odot$ star.

Rogue binary planets and planets orbiting brown-dwarf-candidate
systems have also been detected: MOA-2015-BLG-337 \citep{MOA2015} and
OGLE-2015-BLG-1459L (\citep{hwang2018}2018), the latter potentially
hosting an exomoon. Such systems remain challenging to classify due to
degeneracies, but they confirm microlensing sensitivity to FFPMs.  

Future observatories, such as the Chinese Space Station Telescope
(CSST) and the Nancy Grace Roman Space Telescope, will dedicate time
to microlensing surveys; both are capable of detecting Galilean-mass
exomoons around FFPs. However, simulated light curves indicate they
will recover only a small fraction of systems
\citep{2023MNRAS.520.5613S,roman_moons,CSST_moons}. \citet{roman_moons}
predict $\sim1$ moon detection by Roman, though this depends strongly
on the exomoon mass function, currently unconstrained.

Here, we present N-body simulations used to determine the cross-section for
ejecting a moon together with its planet from their host star due to
the encounter with another star.  In addition, we explore the orbital
parameter space of rogue planet--moon pairs and argue that these help
us to constrain the ejection mechanism.

\section{Methods}\label{sec:methods}
We calculated the cross-section by following the procedure introduced by \citet{HB83} and improved by \citet{mcmillan1996}. N-body simulations were performed using the fourth order predictor-corrector Hermite integrator \citep{hermite} in the AMUSE package  \citep{portegieszwart2009,portegieszwart2013,pelupessy2013,art_of_amuse}.
\subsection{Scattering experiments}
Each scattering experiment starts with a planetary system composed of one $1\, M_\odot$ star orbited by a Jupiter-mass planet and a Ganymede-mass moon. Both planet and moon have co-planar circular orbits with the semi-major axes $a_p$ and $a_m$ for the planet and moon, respectively.

\begin{figure}
    \centering
    \includegraphics[width=\linewidth]{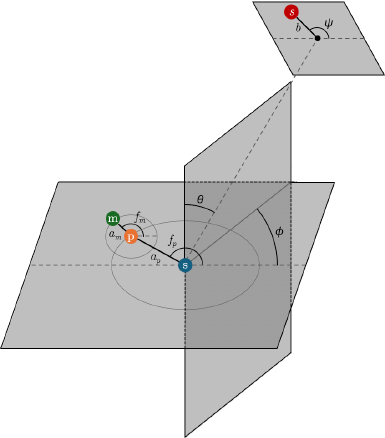}
    \caption{Illustration of initial conditions in each scattering experiment. The solar-mass host star (blue s) is orbited by a Jupiter-mass planet (orange p), which in turn is orbited by a Ganymede-mass moon (green m). The field star (red s) is also of solar mass. The eight independent variables are further explained in Table \ref{tab:vars}. }
    \label{fig:initial_conditions}
\end{figure}

We modeled a close encounter with a field star by injecting a solar-mass intruder into the planetary system. The field star starts at a distance of $20\ a_p$ from the planetary system's center of mass, with initial velocity $v_\infty$ and impact parameter $b$. Its starting position is specified by three angles: $\theta$ and $\phi$ define the direction of approach, and $\psi$ sets the orientation of the orbit in the plane perpendicular to the trajectory. A sketch of the initial geometry is shown in Fig. \ref{fig:initial_conditions}.

The incoming velocity of the field star is always set to $3\,\text{km s}^{-1}$, which is the typical 3D velocity dispersion in a young cluster such as Orion \citep{1988AJ.....95.1755J, 1998ApJ...492..540H}. Note that the planetary system is soft; i.e., its critical velocity is smaller than the cluster's velocity dispersion: $v_\infty>v_\text{crit}$. the two stars cannot form a binary and will always move away from each other after the interaction. As a consequence, the two stars cannot remain bound after the interaction. We were able to stop the simulation once the two stars are unbound, sufficiently far apart, and moving away from each other. We also stopped the simulation if any of the bodies collide with another.

\begin{figure*}
    \centering
    \includegraphics[width=0.49\linewidth]{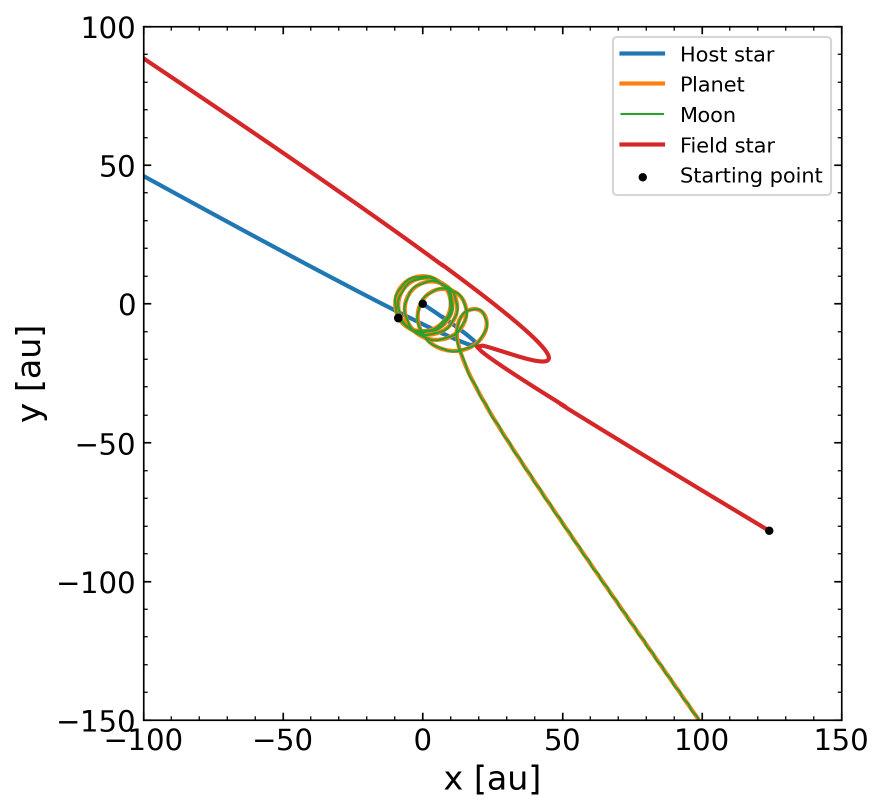}
    \includegraphics[width=0.475\linewidth]{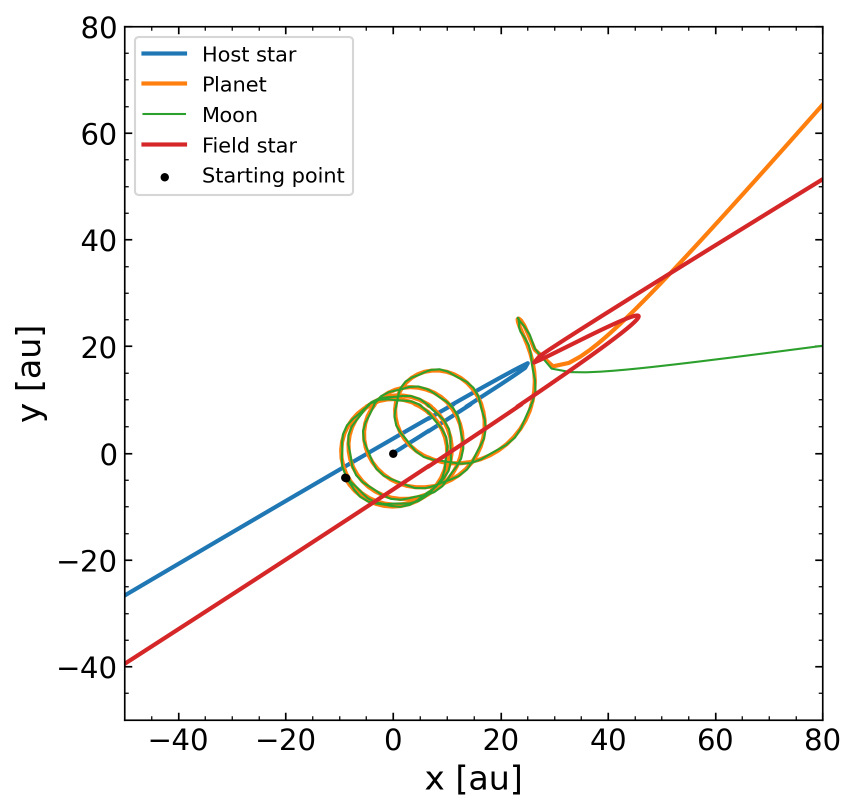}
    \caption{Orbital trajectories of two example scattering experiments. \textit{Left}: Interaction that results in a free-floating planet--moon pair. \textit{Right}: Interaction that results in a free-floating planet and a separate free-floating moon. The starting points of the bodies are indicated by the black dots.}
    \label{fig:spaghetti}
\end{figure*}

In Fig. \ref{fig:spaghetti}, we show two examples for such scattering experiments. The left panel shows the formation of a free-floating planet–moon pair, while in the right panel shows the planet and moon are ejected separately from their host systems, forming two FFPs.

\subsection{Cross-section calculation}

\begin{table*}[]
    \centering
    \caption{Independent variables in the scattering experiments. }
    \begin{tabular}{c|c|c|c}
        \hline
        \hline
        \thead{Parameter} & \thead{Sampling method} & \thead{Probability \\ Homogeneous in} & \thead{Range}\\
        \hline
        $a_p$ & Fixed & --- & ($1$ au -- $100$ au) \\
        $a_m$ & Fixed & --- & ($0.008\,R_H-0.55\,R_H$) \\
        $b$ & Layered & $b^2$ & ($0$ -- Automatically determined)\\
        $\phi$ & Random & $\phi$ & ($0,2\pi$)\\
        $\theta$ & Random & $\cos\theta$ & ($0,\pi/2$)\\
        $\psi$ & Random & $\psi$ & ($0,2\pi$)\\
        $f_p$ & Random & $f_p$ & ($0,2\pi$)\\
        $f_m$ & Random & $f_m$ & ($0,2\pi$)\\
        \hline
    \end{tabular}
    \tablefoot{The second column describes the Monte Carlo sampling method. Column three denotes the variables in which they are uniformly distributed. In the final column, the parameter range is shown.}
    
    \label{tab:vars}
\end{table*}

To determine the cross-section, we ran scattering experiments. We
incrementally increased the impact parameter of the field star until
the desired outcome no longer occurred. We started by running
$n_\text{init}=10$ scattering experiments with impact parameters
between $0$ and $a_p$. We then shift the impact parameter annulus by
$a_p$ each time while increasing the number of simulations $n$ to
maintain constant surface density across all impact parameter ranges.
We performed a total of $33953$ experiments (see
Table\,ref{tab:outcomes}).

The initial conditions for each scattering experiment are sampled in a Monte Carlo fashion, as detailed in Table \ref{tab:vars}. We calculated cross-sections for fixed planet and moon semi-major axes, with the moon semi-major axis expressed as a fraction of the host planet’s Hill radius. This process was repeated for each combination of $a_p$ and $a_m$.

During each cross-section calculation, we recorded the initial conditions and final states of the system. Outcomes are classified based on the final states, allowing us to compute the cross-section
of outcome X using Eq. \ref{eq:cross-section} as a function of the initial semi-major axes of the planet, $a_p$, and the moon, $a_m$:

\begin{align}
    \sigma_X(a_p,a_m) = \pi b^2_\text{max}(a_p,a_m)\frac{n_X(a_p,a_m)}{n_\text{tot}(a_p,a_m)}.
    \label{eq:cross-section}
\end{align}

This method introduces two sources of error. First, statistical uncertainty is quantified using Eq. \ref{eq:stat_err}:
\begin{align}
    \label{eq:stat_err}
    \Delta_\text{stat}\sigma_X(a_p,a_m) &= \frac{\sigma_X(a_p,a_m)}{\sqrt{n_X(a_p,a_m)}}.
\end{align}
Second, some simulations fail due to excessive computation time or high-energy errors, leaving outcomes undetermined. This systematic error is quantified with Eq. \ref{eq:sys_err}:
\begin{align}
    \label{eq:sys_err}
    \Delta_\text{sys}\sigma_X(a_p,a_m)  &= \pi b_\text{max}^2(a_p,a_m)\frac{n_\text{und}(a_p,a_m)}{n_\text{tot}(a_p,a_m)}.
\end{align}

\subsection{Validation}
We validated the two components of the cross-section calculation: individual scattering experiments and initial-condition sampling. Scattering experiments are validated by enforcing a maximum energy error $(<1\%)$ as defined in Eq. \ref{eq:energy_error}:

\begin{align}
    \label{eq:energy_error}
    \frac{\Delta E}{E(0)} = \left|\frac{E(t)-E(0)}{E(0)}\right|.
\end{align}

If the maximum energy error was exceeded, we reran the simulation with
half the internal time step. This process was repeated up to three
times. Simulations failing the energy check even at a minimum time step
are marked as failed. Our code reproduces the results of \citet{HB83}.
In the end, ten simulations (one in 50 000) were discarded because
of too-large energy errors even after this iterative procedure.  Note
that these failed calculations all led to a collision between at least
two objects.

\section{Results}\label{sec:results}
\subsection{Cross-sections}

\begin{figure*}
    \centering
    \includegraphics[width=\linewidth]{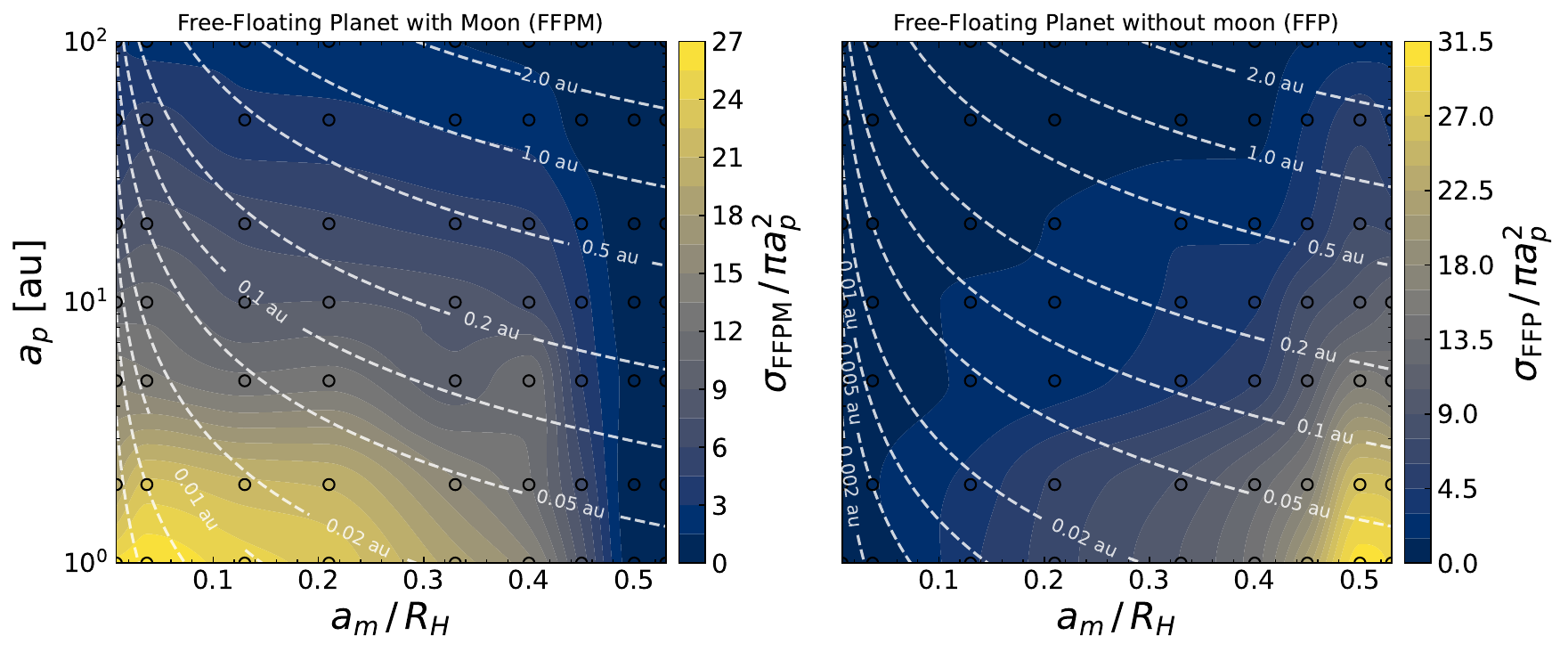}
    \caption{Normalized cross-sections for planet--moon ejections.
      \textit{Left}: Normalized cross-sections as a function of moon and planet semi-major axis for creating a free-floating planet--moon pair. \textit{Right}: Normalized cross-sections as function of moon and planet semi-major axes for creating a free-floating planet without a moon (FFP). The calculated cross-sections (black circles) were interpolated using a PCHIP interpolation scheme. Equal planet-moon separation curves are plotted as dashed white lines.}
    \label{fig:cross-sections_interp}
\end{figure*}

The results of our cross-section calculations are presented in Fig. \ref{fig:cross-sections_interp}. Cross-sections were normalized to the planetary orbital area, $\pi a_p^2$. The moon semi-major axis is expressed as a fraction of the planets' Hill radius ($R_H$). We computed cross-sections on an $a_p$--$a_m$ grid and interpolated between them using a piece-wise cubic hermite interpolating polynomial (PCHIP) scheme \citep{doi:10.1137/0905021}. We focused on interactions resulting in free-floating planets (FFPs without a moon) and free-floating planet--moon pairs (FFPMs).

The normalized FFPM cross-section peaks at the smallest planet and moon semi-major axes, decreasing gradually before a sharp drop at $a_m\sim0.45\, R_H$. For $a_m\lesssim0.45\,R_H$, the cross section is dominated by the planet-moon binding energy, which scales inversely with $a_m$. The drop-off coincides with the external stability radius, $a_E\approx0.48\,R_H$, for pro-grade satellites \citep{domingos2006}. For the closest-in moons, collisions between planets and moons reduce the cross-section.

Figure \ref{fig:cross-sections_interp} (right panel) shows the FFP cross-section, which peaks at $a_m\sim0.5\,R_H$. A sharp increase occurs at $a_m\sim0.45\,R_H$, where moons are either ejected independently or captured by one of the stars, becoming planets.

\begin{figure}
    \centering
    \includegraphics[width=\linewidth]{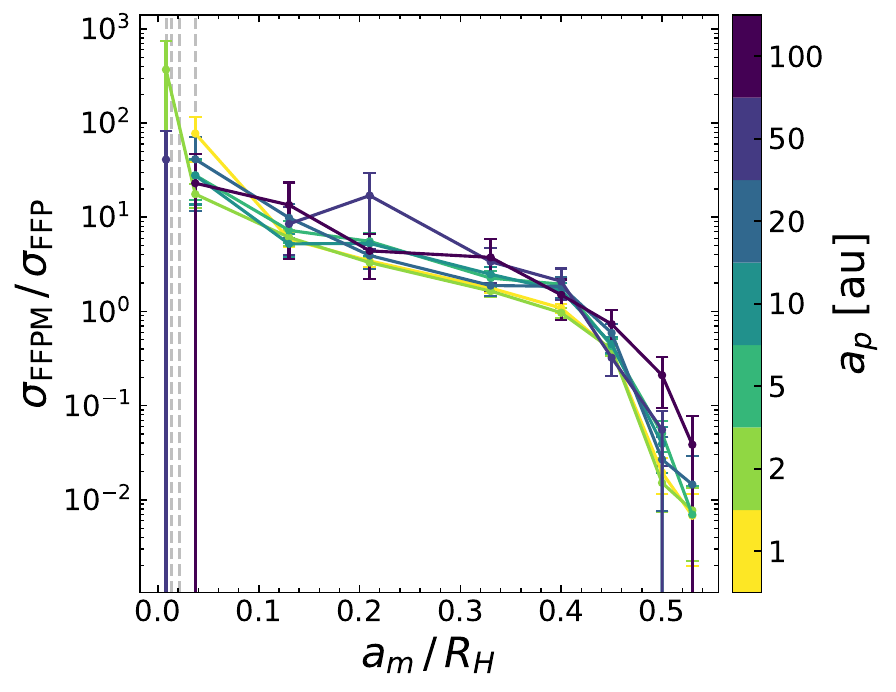}
    \caption{Cross-section ratio between ejecting a planet with and
      without its moon as a function of moon semi-major axis. Colored
      lines indicate the semi-major axis of the planet.  The vertical dashed gray
lines indicate the semi-major axis of Jupiter's
      Galilean moons.}
    \label{fig:cs_ratio}
\end{figure}

Figure \ref{fig:cs_ratio} shows the ratio of cross-sections for creating a free-floating planet with or without a moon as a function, $a_m$. The moon's semi-major axis dominates this ratio, as lines for different planet semi-major axes closely overlap. At a small $a_m$, the FFPM cross-section approaches zero, causing missing values or large error bars in some cases.

\subsection{Moon survival probability}
\begin{figure*}
    \centering
    \includegraphics[width=\linewidth]{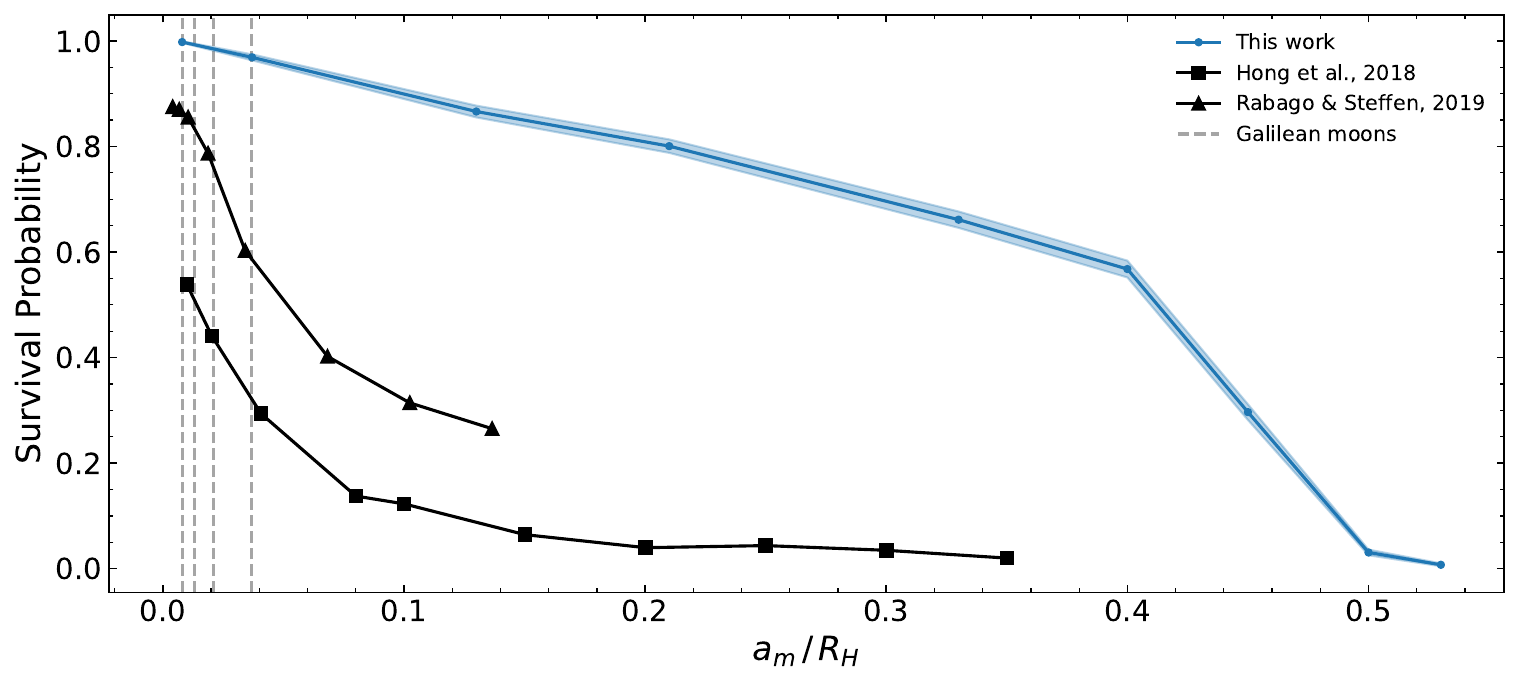}
    \caption{Moon survival probability as a function of initial moon semi-major axis. The one-sigma confidence interval represented by the blue shaded region. We also plot the planet--planet scattering moon survival probabilities from \citet{hong2018} and \citet{rabago2019} to compare the two planet ejection mechanisms. The semi-major axes of the Galilean moons are indicated by dashed gray lines.}
    \label{fig:moon_sma_vs_surv}
\end{figure*}

Figure \ref{fig:moon_sma_vs_surv} shows the survival probability of a moon remaining bound to its host planets during ejection. As expected, the survival probability decreases with increasing planet--moon separation. It declines gradually at first and then drops sharply beyond $0.4\, R_H$, reaching zero at roughly half the planet’s Hill radius. Similar to what is seen in Fig. \ref{fig:cs_ratio}, survival probabilities are nearly identical across planet semi-major axes. Also marked are Jupiter’s Galilean moons, which all have near-unity survival probability. In the hypothetical case that Jupiter is ejected from the Solar System by a stellar encounter, its Galilean moons are
expected to remain bound.

\subsection{Orbital elements}
\begin{figure*}
    \centering
    \includegraphics[width=\linewidth]{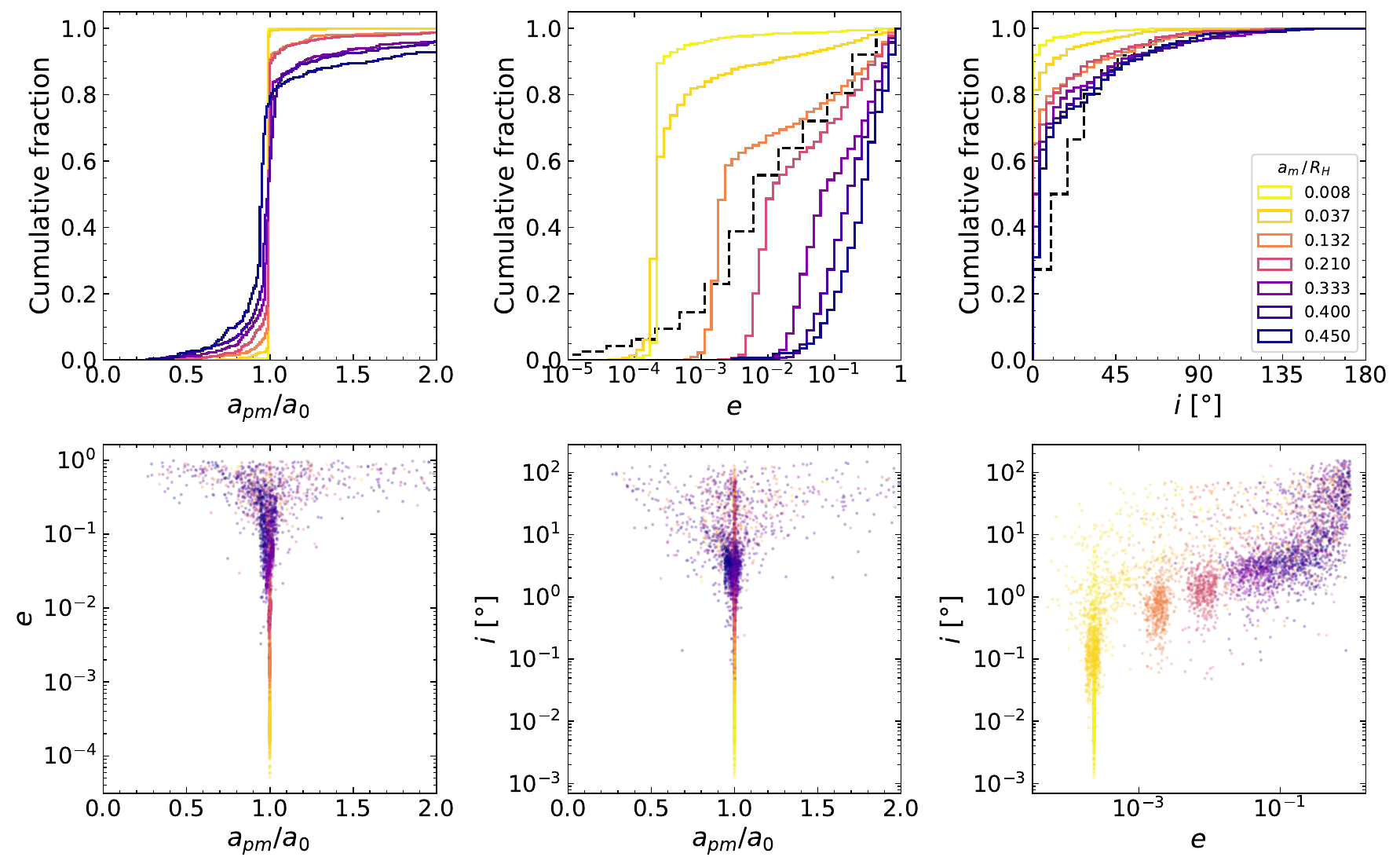}
    \caption{Orbital elements of the created free-floating planet--moon pairs. \textit{Top row}: Cumulative distributions of the change in semi-major axis, eccentricity, and inclination. \textit{Bottom row}: 2D distributions of orbital element combinations. Colors indicate original moon semi-major axis as a fraction of the planets' Hill radius. The orbits of wide moons are perturbed more than the orbits of close-in moons. The dashed black lines show the orbital elements distribution found by \citet{rabago2019}.}
    \label{fig:orb_elements}
\end{figure*}

Figure \ref{fig:orb_elements} shows the distributions of semi-major axis, eccentricity, and inclination for FFPMs formed across all cross-section calculations (we only show the cases for which > 100 FFPMs are produced). Each original moon semi-major axis produces a distinct distribution, with tighter moons less perturbed than wider ones.

The top left panel of Fig. \ref{fig:orb_elements} displays the cumulative distribution of the final planet–moon semi-major axis, $a_m$, relative to the initial $a_0$. Most moons retain their original semi-major axis, particularly tight moons. The fraction experiencing considerable change ($> 0.1 a_0$) increases with larger initial $a_m$.

The moon eccentricity distributions (in log scale) confirms that tight moons remain nearly circular, while wide-orbit moons develop considerable eccentricities. Inclination trends are similar: tight moons retain low inclinations, while wide-orbit moons acquire higher inclinations.

\begin{figure}
    \centering
    \includegraphics[width=\linewidth]{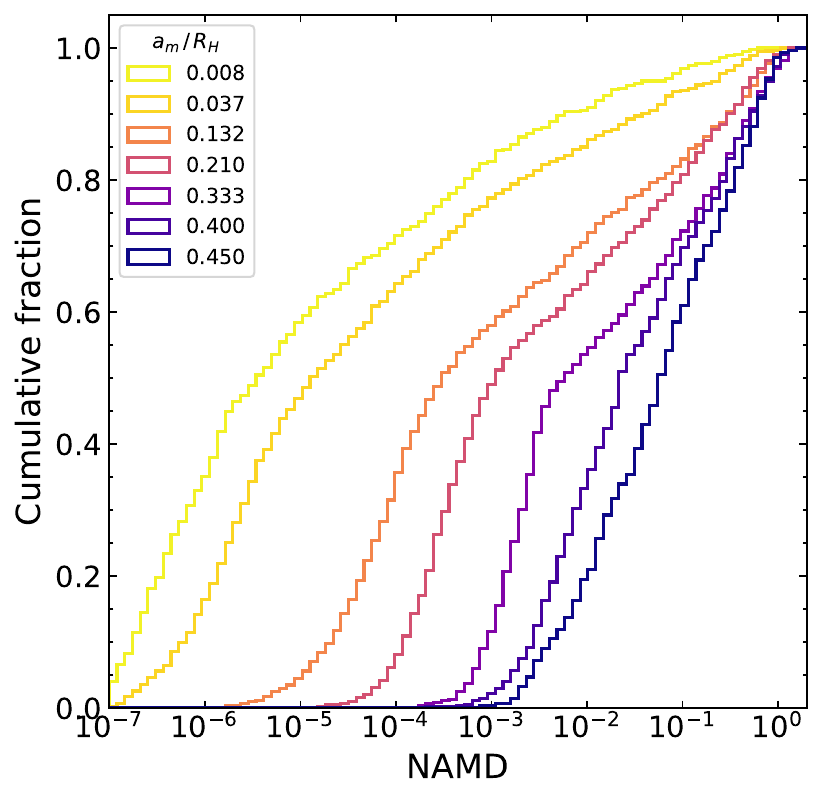}
    \caption{Normalized angular-momentum deficit of rogue planet--moon pairs.
      Cumulative distributions of the NAMD of the formed free-floating planet--moon pairs. Colors indicate the original moon semi-major axis as a fraction of the planets' Hill radius.}
    \label{fig:AMD}
\end{figure}

The bottom left and bottom center panels show eccentricity and inclination versus semi-major-axis change. The bottom right panel demonstrates that wide-orbit moons experience greater orbital
excitation (dynamical heating). Such excited systems have lost angular momentum relatively to their original circular, co-planar configuration (maximum angular momentum). We quantified this via the normalized angular-momentum deficit (NAMD; \citealt{laskar1997,laskar2000,turrini2020}): 
\begin{equation}
    \text{NAMD}=\frac{\sum_k m_k\sqrt{a_k}\left(1-\sqrt{1-e^2_k}\cos i_k\right)}{\sum_km_k\sqrt{a_k}}.
\end{equation}
Which simplifies to Eq. \ref{eq:NAMD_simplified} for a free-floating planet--moon pair:
\begin{equation}
    \label{eq:NAMD_simplified}
    \text{NAMD}=1-\sqrt{1-e^2}\cos i.
\end{equation}
Figure \ref{fig:AMD} shows the NAMD distribution. A NAMD of zero indicates the retention of all angular momentum, while values approaching one reflect losses during the encounter. The NAMD quantifies
the system’s dynamical history \citep{turrini2020}. A small fraction ($1.7\%$) of FFPMs have retrograde orbits ($i > 90\degree$).

\subsection{Retrieving the original planet semi-major axis}
The first panel of Fig. \ref{fig:orb_elements} demonstrates that planet–moon separations
change by less than $10\%$ following $\sim 87\%$ of ejections. Consequently, the observed semi-major axis, $a_{pm}$, of a detected FFPM, when combined with our cross-section calculations, enables us to statistically reconstruct the moon’s original distance to the star within the planet’s Hill sphere.

\begin{figure}
    \centering
    \includegraphics[width=\linewidth]{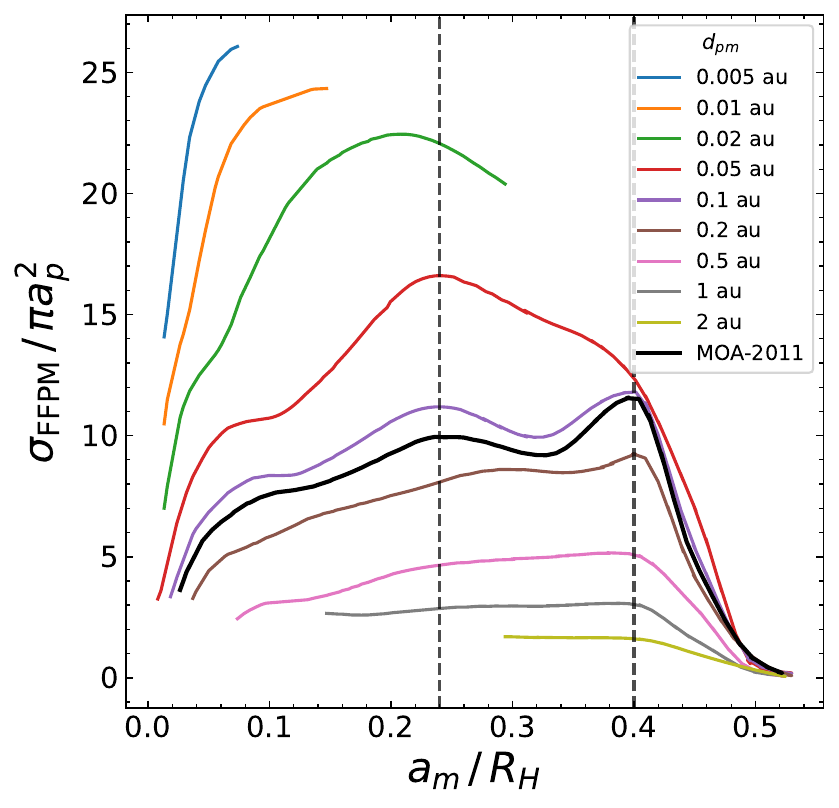}
    \caption{Cross-section for liberating planet--moon pairs.  Rogue
      planet--moon pair formation cross-section as a function of
      initial moon semi-major axis for different planet--moon
      separations, $d_{pm}$. The black curve denotes the cross-section
      curve for the inferred rogue planet--moon interpretation of
      MOA-2011-BLG-262Lb ($d_{pm}\simeq0.13\,\rm au$). Vertical dashed
      black lines indicate the two maxima for this object.}
    \label{fig:original_sma}
\end{figure}

Assuming a host-star mass, this yields the original planet–star separation: $a_p$. Using Fig. \ref{fig:cross-sections_interp}, we present the cross-sections versus initial moon semi-major axis, $a_m$, for various rogue planet–moon separations, $d_{pm}$, in Fig. \ref{fig:original_sma}. The value of $a_m$ at which the cross-section is maximized was then used to estimate the original $a_p$ from before the encounter via Eq. \ref{eq:original_sma}:

\begin{equation}\label{eq:original_sma}
    a_p=\frac{d_{pm}}{a_m/R_H}\left(\frac{m_p}{3M}\right)^{-1/3}.
\end{equation}

Here, $R_H$ is the planet's Hill radius with respect to its original host star of mass $M$, $m_p$ is the mass of the detected rogue planet, $d_{pm}$ is the detected distance between the rogue planet and its moon, and $a_m/R_H$ is the original moon semi-major axis inferred using Fig. \ref{fig:original_sma}.

We applied this method to the candidate rogue planet–moon
MOA-2011-BLG-262Lb \citep{MOA2011} to infer the planet’s original
semi-major axis. The system comprises a planet of mass $m_p \simeq
3.6\,M_\text{Jup}$ and a moon of mass $m_m \simeq 0.54\,M_\oplus$ at
separation $d_{pm} \simeq 0.13\,\text{au}$. The black curve in Fig. 
\ref{fig:original_sma} shows a global maximum at $a_m \simeq 0.4\,R_H$
and a local maximum at $a_m \simeq 0.24\,R_H$. Assuming a solar-mass
host star, this yields original planet semi-major axes in the range of
$a_p \sim 1.3$\,au to $\sim 5.9$, with a slight preference for $a_p
\simeq 5.2\, \text{au}$.

Several assumptions introduce considerable uncertainty, such as the
host-star mass, the moon’s orbital eccentricity (which we assumed to
be circular), and the cross-section analysis, which implicitly assumes a
homogeneous sampling of the initial encounter space.

\subsection{Other outcomes}

There are various outcomes possible from our simulations.  They all
start with a single star that encounters a star orbited by a planet,
which is orbited by a moon, or $(s_1, (p, m)), s_2$.  Such an
experiment can have several outcomes, ranging from preserving the
initial conditions: $(s_1, (p, m)), s_2$, or, as is the focus of our
study is FFPM, which could be written as $s_1, s_2, (p, m)$.  In
Table~\ref{tab:outcomes} we list all possible outcomes for our
experiment, with the relative proportion in which we encountered them.

Because the total energy of our initial systems is positive, the host
star and field star cannot become bound to each other unless the
planet or moon is ejected with unreasonably high velocity (which
did not happen in any of the simulations). We stopped the sampling
procedure when no planet was ejected in a given impact parameter range,
and therefore our data are incomplete in those cases where the planet
has not become free-floating.

\begin{table}[]
\caption{Possible encounter outcomes.
           }
    \centering
        \begin{tabular}{l|l|r}
        \hline
        \hline
        Name & Outcome & \# \\
        \hline
        Initial State        & $(s_1,(p,m)),s_2$ & 10591\\
        Multi-planet system  & $(s_1,p,m),s2$    & 8545\\
        FFPM                 & $s_1,(p,m),s_2$   & 4828\\
        FFP and FFM          & $s_1,s_2,p,m$     & 2315\\
        FFM                  & $(s_1,p),s_2,m$   & 1440\\
        Planet-moon exchange & $s_1,(s_2,(p,m))$ & 1357\\
        FFP                  & $(s_1,m),s_2,p$   & 1008\\
        Multi-planet system  & $s_1,((s_2,p),m)$   & 875\\
        Moon exchange        & $(s_1,p),(s_2,m)$ & 609\\
        FFM                  & $s_1,(s_2,p),m$   & 571\\
        Planet exchange      & $(s_1,m),(s_2,p)$ & 548 \\
        FFP                  & $s_1,(s_2,m),p$   & 496 \\
        Collision            & ---               & 760 \\
        Error too large      & ---               &  10 \\
        \hline
        \end{tabular}
 \tablefoot{Resulting configuration of primary star, $s_1$,
          orbited by a planet--moon pair ($(s_1, (p, m))$) that
          encounters a secondary star, $s_2$.  A bound pair is
          indicated by parentheses.  The majority (10591) of encounters
          resulted in the preservation of the configuration, ten led to
          unrecoverable errors (due to collisions), and the remainder
          classified cases (23352) are listed here. }       
    \label{tab:outcomes}
\end{table}

Note that the free-floating-moon-producing events (4326) are the third
most common result of our adopted encounters, and FFPM events are
the third most common, with a cross-section of about half of the events in
which the moon starts orbiting the parent star rather than the
planet.

\section{Discussion}\label{sec:discussion}
\subsection{Planet-ejection mechanisms}
Close stellar encounters are not the only pathway to ejection. Planet–planet scattering, in which gravitational interactions between giant planets drive one of them out of the system, was also explored by \citet{hong2018} and \citet{rabago2019}. Figure \ref{fig:moon_sma_vs_surv} contrasts moon survival probabilities for these two mechanisms. In our simulations, stellar encounter ejections preserve moons over the full range of semi-major axes, and the resulting
survival is approximately piece-wise linear; this in contrast to the power-law-like survival trends produced by planet–planet scattering.

The two scenarios (planet--planet scattering vs. stellar encounters) predict different survival probabilities at identical orbital separations. \citet{rabago2019} used equal-mass planets, while the mass ratios in \citet{hong2018} span $0.1-1$. \citet{rabago2019} attributed lower survival to increased planet–planet encounters for equal-mass systems, which cumulatively destabilize moon orbits.

The top center and top right panels of Fig. \ref{fig:orb_elements} compare the eccentricity and inclination distributions of FFPMs from planet–planet scattering \citep[][Fig. 3]{rabago2019} with our results. Planet–planet scattering produces more excited orbits: while eccentricities of $e > 10^{-2}$ follow a similar distribution to our case for systems where $a_m/R_H = 0.2$, they lack the characteristic drop-off at lower eccentricities; inclinations, meanwhile, systematically exceed those of even our widest orbit moons $(a_m > 0.15\, R_H)$.

The dominant rogue-planet production mechanism remains debated. Planet–planet scattering alone under-produces observed rogue-planet abundances \citep{2012MNRAS.421L.117V}, as do in situ turbulent fragmentation models \citep{2023Ap&SS.368...17M}. Cluster simulations favor stellar encounters \citep{vanElteren2019}, though planet–planet scattering rates depend strongly on initial system stability, which remains unconstrained.

Moon retention fractions and orbital distributions offer observational discriminants between mechanisms. The future characterization of rogue planet–moon systems could resolve this issue.

\subsection{In situ moon formation around rogue planets}
Several studies found disks around young, free-floating giant planets and brown dwarfs. \citet{2025A&A...704A.170R}, for example, reports infrared excess consistent with circum-planetary disks around young free-floating planets in Upper Scorpius. The authors note a disk fraction of $\sim38\%$ among FFPs. Younger star forming regions show even higher disk fractions \citep{2016ApJ...827...52L,2019AJ....158...54E,2025MNRAS.537.2579S}, indicating that circumplanetary disks around rogue planets are common. In situ moon formation within these disks provides an additional source of exomoons around rogue planets.

Simulating circumplanetary disk/ring survival during planetary ejection would constrain formation histories. The further modeling of moon formation efficiency in surviving disks would quantify relative contributions from distinct rogue planet production mechanisms.

\subsection{Eccentric and inclined orbits}
We restricted parameter space to study circular, co-planar moon orbits. Eccentric and inclined orbits warrant future study. Moons in highly eccentric orbits should be more vulnerable to ejection compared to circular orbits, because they spend more time relatively far from the planet when at apocenter. Mutual inclination also affects survival, as the Hill sphere is not spherical; highly inclined orbits probe closer to this boundary and are therefore prone to being ejected.

\citet{domingos2006} demonstrated that retrograde moons have a larger outer stability radius $(a_E \sim 0.93\, R_H)$ than prograde moons $(a_E \sim 0.4\, R_H)$. However, Solar-System moons with $M > 10^{20}\,\text{kg}$ (except Neptune’s inclined retrograde Triton, $i \approx 156\degree$) exhibit near-zero eccentricities and low inclinations. Thus, eccentric or inclined moons (with respect to the planet’s obliquity) may not represent typical systems, though the Solar System’s representativeness remains questionable.

\subsection{Multi-moon systems}
We simulated systems with a single moon, but multi-moon configurations appear the norm among the Solar System's giant planets, with tens to hundreds of moons. We did not follow multi-moon interactions, and mutual perturbations post-ejection could lead to the subsequent ejection of moons. Our results suggest minimal mutual influence, because the changes in the moon's orbits are small, in particular for the tightest moons (Fig. \ref{fig:orb_elements}).

Mutual influence likely depends on mean-motion resonances (MMRs). \citet{maas2025} demonstrated that resonant planetary systems in star clusters are more stable than non-resonant ones. \citet{rabago2019} found that compact (Galilean-like) lunar MMRs typically survive planet–planet scattering ejections, while wide-orbit MMRs destabilize, losing more moons than test-particle predictions. Close stellar encounters are less strong than planet–planet scattering and lead to more preserved moons. This effect may be enhanced even further when a lunar system is in MMR.

Solar-System planetary-mass moons are tidally locked and experience tidal interactions, which were ignored in our calculations. Tides probably enhance survival for tight orbits, but since those moons are already preserved, tides are not expected to play a prominent role in the survivability of moons orbiting a free-floating planets.

Tidally heated exomoons are observationally accessible \citep{2023A&A...675A..57K,2024A&A...687A.125K,2024A&A...684A..72V}, making post-ejection MMR and heating retention critical for detectability. Tidal interactions also have an impact on rogue exomoon habitability \citep{1997Natur.385..234W,heller2014formation}. In the absence of stellar insolation, tidal heating could lead to subsurface oceans \citep{1987AdSpR...7e.125R} and potentially support life on moons orbiting free-floating planets.

\subsection{Cluster simulations}
Previous simulations modeled entire star clusters with planets
\citep{vanElteren2019} or asteroids \citep{2026A&A...709A..30H}. The
only study on moon-bearing planets in a clustered environment were
performed by \citep{2024ScPA....3....1P}. It turns out that if a
planet--moon pair would be ejected from its parent star while still in
a clustered environment, its survival probability would be low
\citep{2025NatAs...9..957P}. However, such interactions can result in
free-floating binary planets \citep{2024ApJ...970...97Y,2024NatAs...8..756W}, but those studies focused on relatively wide binary planets and planet--moon
pairs, rather than the potential ejection of an entire system of
relatively close moons.

\subsection{Mass ratios}
Moon mass has minimal impact on survival probability, as $a_m$ is normalized to the planet’s Hill radius. The star–planet mass ratio similarly affects both the reference frame and Hill sphere equivalently. The host–field star mass ratio merits investigation, as massive perturbers can eject closer-in planets ($<1\,\text{au}$) via lower $v_\infty$. While absolute cross-sections vary with mass ratios, the qualitative trends identified here should persist.

\section{Conclusions}\label{sec:conclusions}
We conducted numerical scattering experiments to compute cross-sections for planetary ejection with moons during close stellar encounters. Cross-sections for free-floating planet–moon pairs (FFPMs) and free-floating planets without moons (FFPMs) were calculated on an $a_p-a_m$ grid, marginalizing over other initial conditions. Our main findings are listed below.
\begin{itemize}
    \item FFPM cross-sections depend primarily on the moon semi-major axis, $a_m$, with $a_p$ playing a secondary role. Cross sections decrease gradually with $a_m$ and then drop sharply at $a_m\sim 0.45\,R_H$. FFPM cross-sections decrease with increasing $a_p$.
    \item Moon survival probability is near unity for Io-like orbits $(a_m = 0.008\, R_H)$, declining gradually to $0.4\,R_H$. It then drops to zero at $0.5\,R_H$. This contrasts with planet–planet scattering survival \citep{hong2018,rabago2019}, which declines more rapidly from lower initial values. Stellar encounter ejections retain moons more effectively across all $a_m$.
    \item FFPM orbital elements show that tight-orbit moons $(a_m \lesssim 0.4\,R_H)$ retain their inner orbits, while wide-orbit moons are probably excited. Planet–planet scattering FFPMs \citep{rabago2019} lead to higher eccentricities and inclinations.
    \item Observed FFPM separations, $d_{pm}$, statistically constrain the original $a_p$ via cross-section maxima. Applied to MOA-2011-BLG-262Lb \citep{MOA2011}, this yields $a_p = 3.1$ or $5.2 \,\text{au}$ (assuming a solar-mass host), though uncertainties remain substantial.
\end{itemize}

Based on our cross-section calculations, it seems plausible that a rich system of moons survives the ejection of its host planet, giving rise to free-floating systems. So far, no such systems have been confirmed, although we remain quite intrigued by the candidate planet-moon MOA-2011-BLG-262Lb and the disks found around the rogue planets MOA-2015-BLG-337 and OGLE-2015-BLG-1459L.

We do not know, however, how frequently rogue planets have moons, and if their existence is confirmed we do not know their ejection-survival or formation mechanism. Moon-retention fractions and orbital properties offer potential discriminants to answer this question, although it seems difficult to confirm their origin.

\begin{acknowledgements} 
It is a pleasure to thank Erwan Hochart for suggesting the study of rogue moons orbiting rogue planets, and for participating in the initial discussions that led to this work.This work was performed using the compute resources from the Academic Leiden Interdisciplinary Cluster Environment (ALICE) provided by Leiden University.
\end{acknowledgements}

\bibliographystyle{aa}
\bibliography{refs.bib}

\end{document}